\documentclass{pic2012}
\def\runauthor{Yvonne Peters}
\def\shorttitle{$t\bar{t}$ Asymmetries at the Tevatron and the LHC }

\newcommand{\met}       {\mbox{$\not\!\!E_T$}}
\newcommand{\mcatnlo}    {\mbox{\textsc{mc@nlo}}}

\begin{document}

\title{top anti-top  Asymmetries at the Tevatron and the LHC}

\author{Yvonne Peters, \\ on behalf of  the ATLAS, CDF, CMS and D0 Collaborations}

\address{University of G\"ottingen, also at DESY\\
Friedrich-Hund Platz 1, 37077 G\"ottingen, Germany\\
E-mail: reinhild.peters@cern.ch }

\maketitle
\abstracts{The heaviest known elementary particle today, the top quark,
  has been discovered in 1995 by the CDF and D0 collaborations at the
  Tevatron proton antiproton collider at Fermilab. Recently, the CDF and D0
  collaborations have studied the forward-backward asymmetry in
  $t\bar{t}$ events, resulting in measured values larger than the standard
  model prediction. With the start of the LHC at CERN in
  2010, a new top quark factory has opened and asymmetry measurements
  in $t\bar{t}$ have also been performed in a proton proton
  environment with higher collision energy. No deviations from the standard model have
  been noticed so far in the measurements of ATLAS and CMS. This article discusses recent
  results of asymmetry measurements in $t\bar{t}$ events of the ATLAS,
  CDF, CMS and D0 collaborations. }

\section{Introduction}
The top quark was discovered in 1995 in proton anti-proton collisions
at a centre-of-mass energy of 1.8 TeV by the CDF and D0 collaborations
at the Tevatron~\cite{cdfdiscovery,d0discovery},
and is the heaviest known elementary particle today.  Due to its high  mass of $m_t = 173.18
\pm 0.94$~GeV~\cite{topmass} and its short lifetime, the top quark is believed to play a special role in
electroweak symmetry breaking, serves as a window to physics beyond the
standard model (SM), and provides a unique environment to study a bare
quark.

As of today, two colliders with high enough energy exist or did exist  where top
quarks can be produced.  Top quarks are produced at the Tevatron $p\bar{p}$ collider with
a centre-of-mass energy of 1.96~TeV, that operated until September
30th 2011, and at the Large Hadron Collider (LHC)
at CERN, colliding protons on protons  with centre-of-mass energies of
7~TeV (2011 data) and 8~TeV (2012 data). Due to the high
centre-of-mass energy, the top antitop quark pair ($t\bar{t}$)
production cross section at LHC is approximately 20 times larger than at the
Tevatron~\cite{andreas}. Furthermore, the main $t\bar{t}$  production
process is via gluon-gluon fusion at the LHC.

In this article, analyses of $t\bar{t}$ asymmetries are presented as performed by the
CDF and D0 experiments at the Tevatron at 
Fermilab and the ATLAS~\cite{atlas} and CMS~\cite{cms} experiments at the LHC at CERN. The
results are based on up to 8.7~fb$^{-1}$ of $p\bar{p}$ collision data
for the Tevatron experiments and up to 5.0~fb$^{-1}$ of $pp$ collision
data taken during the 7~TeV run of the LHC experiments in 2011. The results
were obtained in the dileptonic and in the lepton plus jets $t\bar{t}$
final state ($\ell$+jets). Details about the $t\bar{t}$ production and
the classification into different channels are given in Ref.~\cite{andreas}. While at the Tevatron $t\bar{t}$
asymmetries larger than the SM predictions have been observed, the
measured values and the theory predictions of the charge asymmetry at
the LHC are in good agreement so far. In addition to the inclusive
asymmetries, the $t\bar{t}$ asymmetries have been studied as function
of several variables, showing an enhanced dependency at the Tevatron
compared to the SM prediction. Details about the individual
results at Tevatron and LHC are provided in the following sections.

\section{Asymmetry Definitions}
At leading order (LO) quantum chromodynamics (QCD), the production 
of $t\bar{t}$ events is forward-backward symmetric in quark antiquark
annihilation processes. However, at higher order calculations,
intereferences between different diagrams cause a preferred direction
of the top quark and the antitop quark and thus an asymmetry. 
In particular, at next to leading order (NLO) QCD, the leading
contribution to the asymmetry arises from the interference between
tree-level and box diagrams, resulting in a positive asymmetry with
the top quark preferentially being emitted in the direction of the
incoming quark. In addition to the dominant contributions from quark
antiquark annihilation, the process with a quark and a gluon in the
initial state also contributes to the $t\bar{t}$ asymmetry. 

At the Tevatron, which is a $p\bar{p}$ collider,  the $t\bar{t}$
production is dominated by the interaction of a valence quark and a
valence antiquark. Therefore,  the quark
direction can be assumed to  coincide with the direction of the incoming proton, and the
antiquark direction with the incoming antiproton. 
The forward backward asymmetry can be defined in terms of the difference
between the rapidity of the top and antitop quarks, $\Delta y = y_t - y_{\bar{t}}$, as 
\begin{equation}
A_{fb}^{t\bar{t}}=[N(\Delta y >0) - N(\Delta y<0)]/[N(\Delta y >0) +                                                                                                                                                                                                                  
N(\Delta y<0)],
\end{equation}
 where $N(\Delta y >0)$ and $N(\Delta y <0)$ are the
number of events with rapidity difference larger and smaller zero,
respectively.

At the LHC, which is a $pp$ collider, the measurement of the asymmetry
is more challenging for two reasons. Firstly, at $\sqrt{s}=7$~TeV, the $t\bar{t}$ production is dominated by gluon-gluon fusion,
which contributes about 85\% to the total $t\bar{t}$ production cross-section. The gluon-gluon fusion process does not contribute to the
$t\bar{t}$ asymmetry. Secondly, the direction of the incoming quark is
not known due to the collision of two protons. The asymmetry definition
used for the measurements performed by the ATLAS and CMS
collaborations relies on the fact that $t\bar{t}$ production via
$q\bar{q}$ annihilation is dominated by valence quarks, which carry a
large momentum fraction, and antiquarks from the sea, having a smaller
momentum fraction on average. An asymmetry, where the top quark is
preferentially emitted into the direction of the incoming quark thus
results in a wider rapidity distribution for the top quarks compared to
the antitop quarks. The asymmetry measurements at ATLAS and CMS are
therefore performed using the charge asymmetry 
\begin{equation}
A_{C}=[N(\Delta |y| >0) - N(\Delta |y|<0)]/[N(\Delta |y| >0) +                                                                                                                                                   
N(\Delta |y|<0)], 
\end{equation}
 where $\Delta |y|$ is the difference of the
absolute rapidity of the top and antitop quark.

In addition to these definitions, the asymmetry can also be extracted
using the rapidity of the leptons only, the rapidity difference of the
leptons or the difference in number of events with rapidity of the
lepton different from the antilepton,
\begin{eqnarray}
A_{fb}^{l} & = & [N(q_l  y_l >0) - N(q_l y_l<0)]/[N(q_l y_l >0) +                                                                                                                                                                                                                  
N(q_l y_l<0)] \\
A^{ll}_{fb} & = & [N(\Delta \eta >0) - N(\Delta \eta<0)]/[N(\Delta \eta >0) +                                                                                                                                                                                                                  
N(\Delta \eta<0)]\\
A^{CP}_{fb} & = & [N_{l^{+}}(\eta >0) - N_{l^{-}}(\eta
<0)]/[N_{l^{+}}(\eta >0) + N_{l^{-}}(\eta <0) ] \\
A^{ll}_C & = & [N(\Delta |\eta| >0) - N(\Delta |\eta|<0)]/[N(\Delta |\eta| >0) +                                                                                                                                                   
N(\Delta |\eta|<0)],
\end{eqnarray}
where $q_l$ is the charge of the lepton and $y_l$ is the rapidity of the
lepton and $\Delta \eta$ is the pseudo-rapidity difference of the
lepton and antilepton. $N_{l^{+}}(\eta >0)$ and $N_{l^{-}}(\eta <0)$
are the number of events with the antilepton having positive
pseudorapidity and the lepton negative pseudorapidity, respectively. 
The advantage of the asymmetry measurements which use leptons is that
the complete reconstruction of the $t\bar{t}$ system is not
necessary.  Furthermore, the leptonic asymmetry provides additional
information with respect to the forward-backward asymmetry, since it is sensitive to
polarization effects.

\section{$t\bar{t}$ Asymmetries at the Tevatron}
The first time the $t\bar{t}$ forward-backward asymmetry has been
measured was by the CDF and D0
collaborations~\cite{cdffirstasym,d0firstasym}, where both measure values
larger than the SM prediction. In this section,  a short overview
over SM calculations of the $t\bar{t}$ asymmetry is given, followed by
recent results from the CDF and D0 experiments are discussed. 

\subsection{Theoretical Predictions}
The $t\bar{t}$ asymmetry is zero at LO QCD, but larger than zero in NLO QCD
calculations. Therefore, any theoretical NLO QCD $t\bar{t}$ calculation only
yields a LO calculation of the $t\bar{t}$ forward-backward asymmetry.
Besides NLO QCD calculations, several
calculations have been performed that  include additional effects or higher
order diagrams for the $t\bar{t}$
asymmetry at the Tevatron. At NLO QCD, the $t\bar{t}$ forward-backward
asymmetry is predicted to be $A_{fb}^{t\bar{t}}=
7.32^{+0.69}_{-0.59} {\phantom{i}}^{+0.18}_{-0.16}$\%~\cite{kuhnrodrigo,pecjak}, where the
uncertainties include uncertainties due to factorization and
renormalization scale variations as well as uncertainties on the
choice of parton distribution function (PDF). 
Calculations including next-to-next-to-leading-logarithmic (NNLL)
contributions (NLO+NNLL) ~\cite{ahrens} or
calculations at approximate next-to-next-to-leading order (NNLO$_{\rm
  approx}$)~\cite{kidonakis} have been performed, finding
$A_{fb}^{t\bar{t}}= 7.24^{+1.04}_{-0.67} {\phantom{i}}^{+0.20}_{-0.27}$\% for the
NLO+NNLL calculation ~\cite{ahrens}. Additionally, calculations including 
effects due to electroweak and mixed QCD electroweak corrections
have been performed. For example inculding contributions from
$b\bar{b} \rightarrow t\bar{t}$ diagrams~\cite{bernreuther2010}, changing
the asymmetry by a relative amount of 5\% with respect to the NLO QCD calculation, or
including effects from photonic corrections~\cite{holikpagani}, which
enhances the asymmetry by a relative amount of 22\% compared to the
NLO QCD calculation. A recent calculation including  electroweak and
mixed QCD and electroweak corrections as well as using a NLO PDF in
the denominator of the expansion in $alpha$ and $alpha_S$ yields an asymmetry of
$A_{fb}^{t\bar{t}}= 8.8\pm 0.6$\%~\cite{bernreuther2012}. In a similar
calculation performed recently, similar effects have been
observed~\cite{kuhnrodrigo2012}. A more detailed overview over
theoretical predictions of the $t\bar{t}$ asymmetry and the included
corrections can be found for example in Ref.~\cite{pecjak}. Even though
there are arguments that the asymmetry value should not change much at
higher order calculations, there are no calculations at full NNLO QCD
available as of today. The different calculations presented here show
an enhancement of the $t\bar{t}$ asymmetry of about 20\% or more when
including additional corrections with respect to the NLO QCD
calculation. When comparing the experimentally measured asymmetry
value to the theoretical prediction, this can have a sizeable effect
on the deviation of the  measurement and the theoretical prediction. 

\subsection{Asymmetry Measurements}
The CDF and D0 collaborations have performed $t\bar{t}$ asymmetry
measurements in the $\ell$+jets and dilepton final
states. These analyses require one or both of the $W$ bosons from the
top quark to decay leptonically. In the
$\ell$+jets final state, exactly one isolated, high $p_T$ electron or
muon, large missing transverse energy (\met) due to the undetected neutrino from the $W$ boson decay, and four or more jets with large
transverse momentum $p_T$ are required. The main background contributions in the
$\ell$+jets final state consist of $W$+jets production and
instrumental background due to QCD-multijet events in which jets are misidentified as leptons. 
Additional selection cuts were introduced to
reduce the instrumental background, as for example on the azimuthal angle
between the lepton momentum and the direction of the missing
transverse energy. At least one of the jets is required to be
identified as a $b$-jet to enhance the purity of the sample. 
The signature in the dilepton final state consists
of two isolated, high-$p_T$ leptons ($ee$, $e\mu$ or $\mu\mu$), at
least two high $p_T$ jets and large \met\ from the
two neutrinos. 

In order to measure the forward-backward $t\bar{t}$ asymmetry, the
reconstruction of the full $t\bar{t}$ event is required. A kinematic
fitter was used for this purpose, which include constraints from the
known 
$W$~boson mass and top quark mass to extract the missing information
about the neutrino momentum and obtain the jet combinations matching the
top and the antitop quarks. The CDF collaboration also extract
$A_{fb}^{t\bar{t}}$ in the dilepton final state, necessitating the
full $t\bar{t}$ reconstruction of the dileptonic events. For this
purpose, an algorithm was used, which compares calculated longitudinal
and transverse momenta of the $t\bar{t}$ system as well as the
invariant $t\bar{t}$ mass to probability distribution functions 
of these variables based on standard model expectations. The most
likely solution was then chosen using a likelihood function based on
these probability density functions. 

In the $\ell$+jets final state, the background determination in the
analysis performed by D0 was carried out by fitting a topological likelihood
function, based on variables that are uncorrelated to $\Delta y$. In the analysis performed by the CDF collaboration, the background
was estimated using data-driven methods in samples orthogonal to the
signal sample and using Monte Carlo (MC) predictions. 
From the signal samples, the distributions of the lepton rapidity and
$\Delta y$ were extracted and the background distributions were
subtracted from the data. Up to  this step in the analysis chain, the extraction of the
rapidity and $\Delta y$ distributions is mostly independent of the
modelling of the signal. Due to
acceptance effects and detector resolutions, the asymmetry results
extracted from these distributions can not be compared to theoretical
predictions or between the experiments. In order to correct for these
effects, both CDF and D0 apply unfolding techniques on the rapidity or
$\Delta y$ distributions. In the analysis performed by the CDF
collaboration a $4\times 4$ matrix inversion is applied on the $\Delta
y$ distribution, while at D0 regularized unfolding has been used. 

After unfolding, an inclusive asymmetry of $A_{fb}^{t\bar{t}}=16.2 \pm4.2$\% has
been extracted by CDF using 8.7~fb$^{-1}$ of Tevatron Run~II
data~\cite{cdfljetsnew}, and $A_{fb}^{t\bar{t}}=19.6 \pm 6.5$\% by D0
using 5.4~fb$^{-1}$ of data~\cite{d0ljets}, where both results are limited
by statistical uncertainties. Comparing these results to a NLO
prediction obtained by using MC,  corrected for electroweak and QCD effects, of
$A_{fb}^{t\bar{t}}=6.6$\%~\cite{cdfljetsnew}  implies  that both
  measurements are about two standard deviations higher than the SM
  value. Both CDF and D0 performed several studies on potential
  influences from signal or background modelling on the measurement, as
  for example checking the modelling of the asymmetry in $W$+jets
  events using events with no identified $b$-jets or checking the
  dependence of the asymmetry in the MC simulation on the transverse
  momentum of the $t\bar{t}$ system, $p_T^{t\bar{t}}$. The
  latter study showed that colour coherence effects in the MC
  simulation can
  introduce an asymmetry depending on $p_T^{t\bar{t}}$ even in LO
  MC. This effect is included in the systematic uncertainties. 

Besides the measurement of $A_{fb}^{t\bar{t}}$, the lepton-based
asymmetry $A_{fb}^{l}$ has been extracted by both
collaborations in the $\ell$+jets final state. Since the resolution of the lepton rapidity is very
good, the unfolding in this measurement is much simpler than what is
needed for $A_{fb}^{t\bar{t}}$, and no reconstruction of the
$t\bar{t}$ system is required. Using 5.4~fb$^{-1}$, D0 extracts
$A_{fb}^{l}= 14.2 \pm 3.8$\% with $|y_l|<1.5$, and CDF measures $A_{fb}^{l}=6.6 \pm
2.5$\% using 8.7~fb$^{-1}$. The D0 collaboration compared this
measurment to the prediction using \mcatnlo\ MC~\cite{mcnlo}, which yields
$A_{fb}^{l}=0.8$\%. Therefore, the measurement is more than three
standard deviations higher than this prediction. The CDF collaboration
compared their result to a NLO
prediction including electroweak and QCD effects,
$A_{fb}^{l}=1.6$\%~\cite{bernreuther2012}. 

\begin{figure*}[t]
\centering
\includegraphics[scale=0.20]{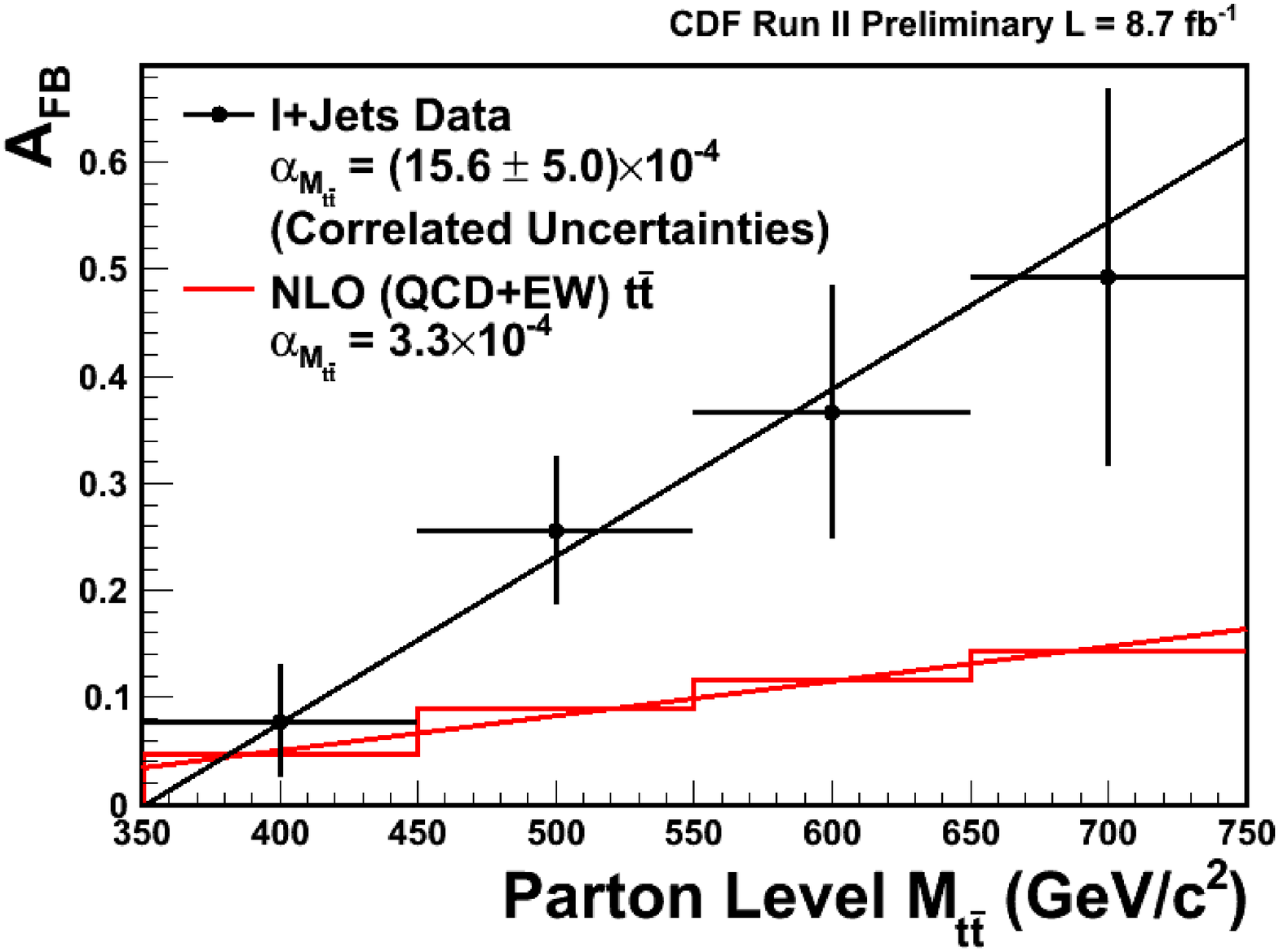}
\includegraphics[scale=0.20]{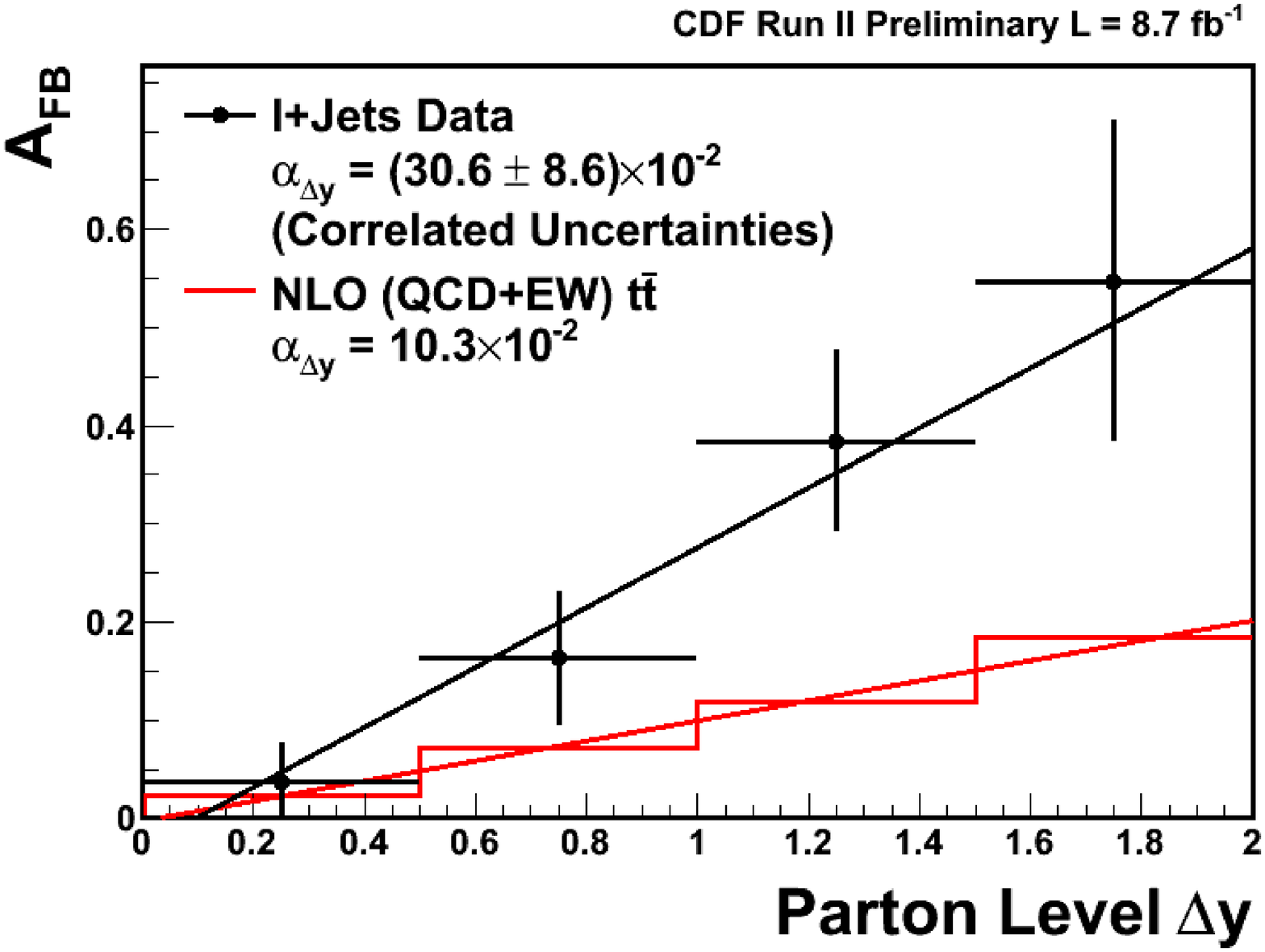}
\caption{Parton level $A_{fb}^{t\bar{t}}$  as  function of
  $m_{t\bar{t}}$ (left) and $\Delta y$ (right) as extracted by the CDF
  collaboration in the full Run~II data set. The best-fit line is
  superimposed \protect \cite{cdfljetsnew}. } \label{massdep}
\end{figure*}

The asymmetry is expected to depend on several variables, as for
example on the invariant $t\bar{t}$
mass, $m_{t\bar{t}}$, the rapidity $y_t$ and $p_T^{t\bar{t}}$. For example, a
dependency of the $t\bar{t}$ asymmetry on $m_{t\bar{t}}$ is expected, since the relative fraction of
$t\bar{t}$ production due to quark antiquark annihilation is enhanced
with increasing $m_{t\bar{t}}$.
Besides the inclusive measurements, both collaborations also measured
the asymmetry as a function of $m_{t\bar{t}}$ and $\Delta y$. Both collaborations noted a
stronger dependency of $A_{fb}^{t\bar{t}}$ on $\Delta y$ and
$m_{t\bar{t}}$ than predicted by the SM. In particular,  the
asymmetry measured by CDF for $m_{t\bar{t}}>450$~GeV deviates from the
prediction by more than two standard deviations. In Fig.~\ref{massdep}
the parton-level asymmetry as function of $m_{t\bar{t}}$ (left) and
$\Delta y$ (right) are shown when using four bins in $m_{t\bar{t}}$ or
$\Delta y$. Very recently, the CDF collaboration updated their
measurement in the $\ell$+jets final state to the full Tevatron Run~II
data set of 9.4~fb$^{-1}$~\cite{cdfljetssupernew}.



In the dilepton final state, several asymmetries have been explored by
the two collaborations. The CDF collaboration extracts
$A_{fb}^{t\bar{t}}$ and $A_{fb}^{l}$, while the analysis performed by
the D0 collaboration concentrates on lepton-based asymmetries, namely 
$A_{fb}^{l}$, $A^{ll}_{fb}$, and $A^{CP}_{fb}$. Using 5.1~fb$^{-1}$ of
data, CDF measures $A_{fb}^{t\bar{t}}=42 \pm 16$\%, with a prediction
of $6 \pm 1$\%~\cite{cdfdilepton}, and  $A_{fb}^{l} =14\pm 5$\%, where
the latter result is without corrections for acceptance or resolution
effects. The results extracted by D0 are based on 5.4~fb$^{-1}$ of
Run~II data~\cite{d0dilepton}, yielding $A_{fb}^{l}= 5.8 \pm 5.3$\% with a prediction of
$4.7\pm0.1$\%, $A^{ll}_{fb}=5.3 \pm 8.4$\% comparable to a prediction
of $6.2\pm 0.2$\%, and $A^{CP}_{fb} = -1.8 \pm 5.3$\% with a
prediction of $-0.3 \pm 0.1$\%. The results from D0 are after
corrections for acceptance and resolution effects. In addition, D0
performed a combination of the $A_{fb}^{l}$ measurement in the
$\ell$+jets and dilepton final state, yielding $A_{fb}^{l}=11.8 \pm 3.2$\%~\cite{d0dilepton}.

\begin{figure*}[t]
\centering
\includegraphics[scale=0.31]{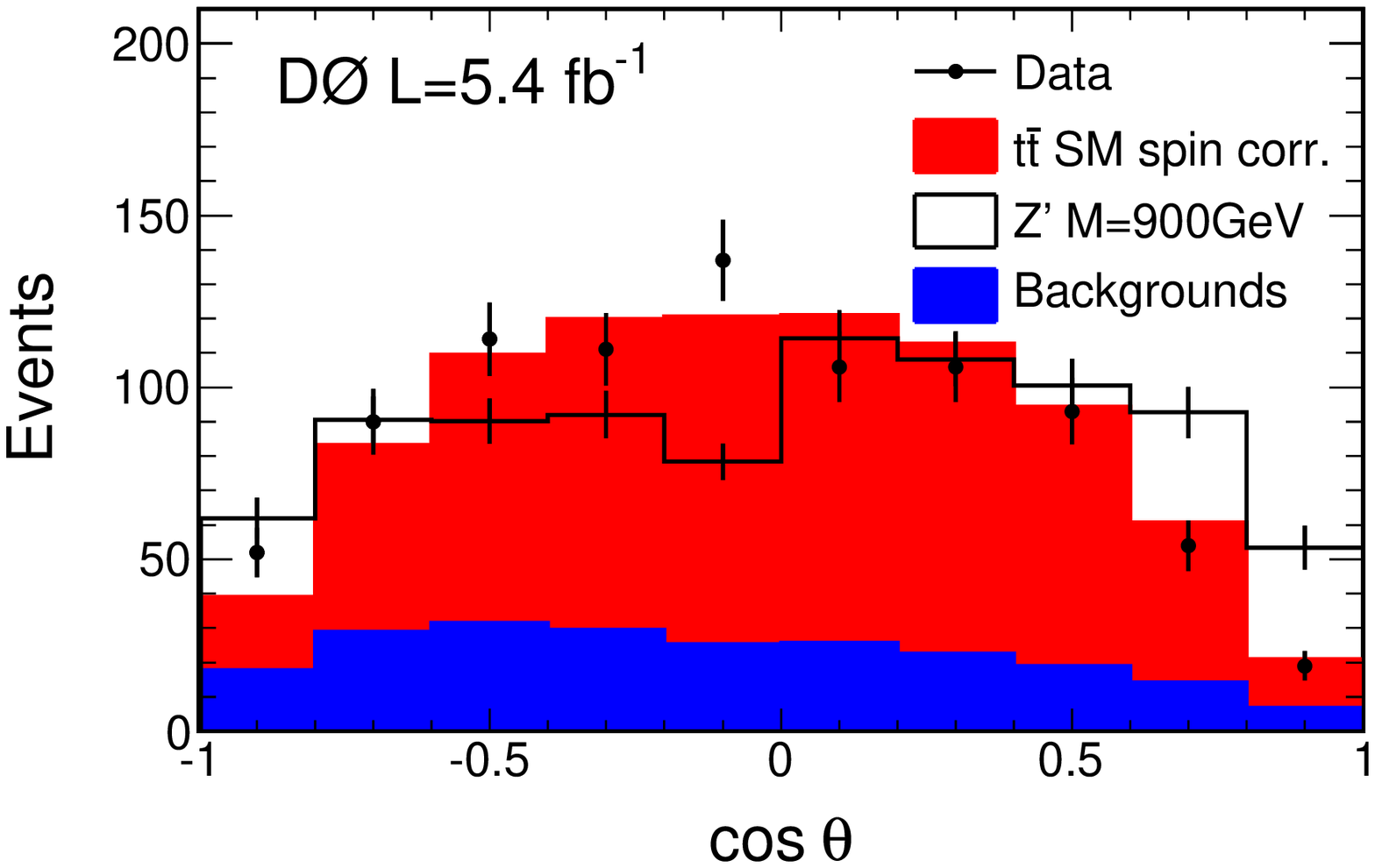}
\includegraphics[scale=0.30]{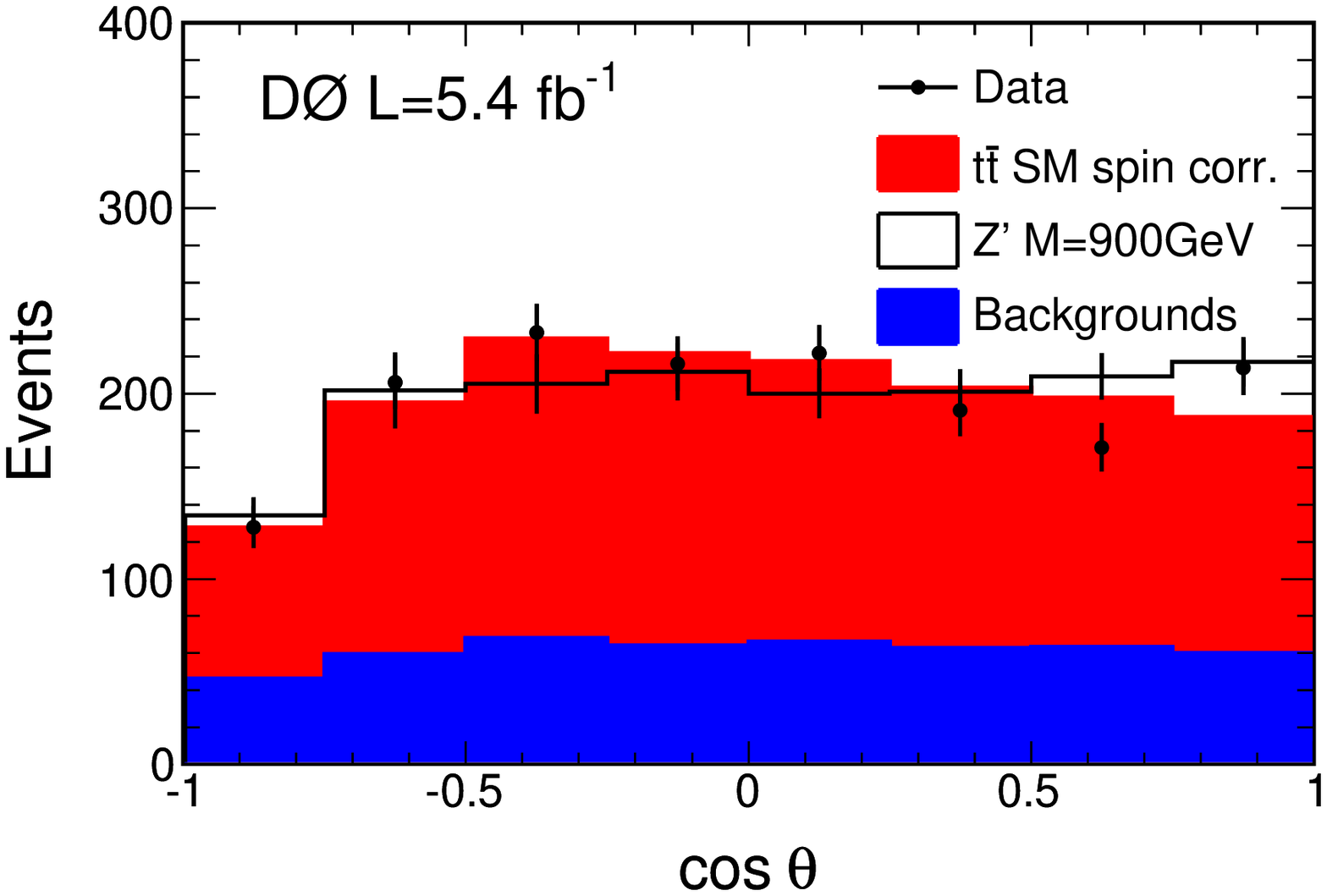}
\caption{The distribution of $\cos \theta$ is  shown for the
  combination of the dilepton channels (left) and the $\ell$+jets
  channels (right). The data are compared to the SM predictions and to
  a model including a hypothetical $Z'$ boson is used \protect \cite{d0dilepton}. }  \label{polarization} 
\end{figure*}

Since the measurements of the $t\bar{t}$ asymmetry performed by CDF
and D0 show a deviation from the SM prediction of two standard
deviations and more, a large interest arose in the theory community,
resulting in an influx of models beyond the SM that could explain the
large positive asymmetries. For an overview of many of these models,
see for example Ref.~\cite{bsmmodeloverview}. Most of these models
already have to fulfill several constraints based on existing top quark
production and properties measurements. For example, the models are
constrained by the observed $t\bar{t}$  production cross-section, and no
significant same-sign top quark pair production should be introduced
since tight limits have been set in direct searches. In order to
distinguish between the models, 
additional measurements have to be performed. One such measurement is
the study of top quark polarization, defined by 
\begin{equation}
\frac {1} {\Gamma}  \frac {d\Gamma} {d\cos \theta_{i,n}} = \frac{1}{2}
(1 + P_n \kappa_i \cos \theta_{i,n} ) \,
\end{equation}
where $\Gamma$ is the decay width, $P_n$ is the polarization,
$\kappa_i$ is
the spin analysing power of the decay product $i$ and $\theta_{i,n}$
is the angle of the decay product $i$ with respect to a chosen
quantization axis~\cite{spincorr,polarizedview}. 
In the SM, the top-quark
polarization at hadron colliders is negligible. Many of the models predicting a
positive $t\bar{t}$ asymmetry also predict a top quark polarization
significantly different from zero, for example due to a
new parity-violating interaction that affects the $t\bar{t}$ production and
leads to a longitudinal polarization of the top quark.
The D0 collaboration performed the first study of the top quark
polarization in the dilepton and $\ell$+jets final state, using
5.4~fb$^{-1}$ of data~\cite{d0dilepton}. The required reconstruction
of the $t\bar{t}$ system in the dilepton final state has been
performed using the neutrino weighting
technique. The $\cos \theta$ distribution is constructed using the
charged leptons, as their spin analysing power is one at LO. As quantization
axis the helicity basis is used.  Figure~\ref{polarization} shows the distribution of $\cos
\theta$ for the dilepton (left) and $\ell$+jets (right) final state,
compared to the SM prediction and a hypothetical $Z'$ boson~\cite{Harris1999ya}. The agreement
between the SM predictions and the data is good for both channels.

\section{$t\bar{t}$ Asymmetries at the LHC}
The ATLAS and CMS collaborations have performed several asymmetry
measurements in the $\ell$+jets and the diletpon final state. 
In particular, the ATLAS collaboration has extracted a result for
$A_C$ inclusively and as function of $m_{t\bar{t}}$  in the
$\ell$+jets final state using 1.04~fb$^{-1}$ of $pp$ collision data
with 7~TeV centre-of-mass energy~\cite{atlasljets}, and has measured
$A_C$ and $A_C^{ll}$ in the dilepton final state using the full 7~TeV
data set of 4.7~fb$^{-1}$~\cite{atlasdilep}. The CMS collaboration has
measured $A_C$ in the $\ell$+jets final state using the full 7~TeV
data sample of 5.0~fb$^{-1}$, where the asymmetry has been measured as
 function of $m_{t\bar{t}}$, the rapidity of the top and
$p_T^{t\bar{t}}$ and inclusively~\cite{cmsljets}. 

The principle of the asymmetry measurements is similar to that for the
measurements at the Tevatron: After selecting a signal
sample~\cite{andreas}, the $t\bar{t}$ final state is
reconstructed and  the distribution of the absolute rapidity of
the top and the antitop quarks are measured. The $t\bar{t}$
reconstruction is carried out using a kinematic fitter. For the $A_C^{ll}$ measurement, the
reconstruction of the $t\bar{t}$ system is not necessary, and the
distribution of pseudorapidity of the two  leptons is studied
instead. To correct for acceptance and detector effects, unfolding of
the distribution is performed.

In the $\ell$+jets final state, the $\Delta |y|$ distributions are
unfolded using iterative Bayesian unfolding by ATLAS, while the CMS
collaboration used a regularized unfolding technique. The ATLAS
collaboration extracts a value of $A_C=-1.9 \pm 2.8 {\rm (stat)}
\pm 2.4{\rm (syst)}$\% in the 1.04~fb$^{-1}$ data sample, while the
prediction with \mcatnlo\ is $A_C=0.6 \pm 0.2$\%. The value extracted
by the CMS collaboration is $A_C=0.4 \pm 1.0 {\rm (stat)} \pm 1.1 {\rm
  (syst)}$\% on 5.0~fb$^{-1}$ of data. For both measurements, the
  systematic uncertainties are comparable in size to the statistical
  uncertainties. The dominant systematic uncertainties are related to
  the modelling of $t\bar{t}$ events, to the uncertainty on the jet
  energy scale, and to the unfolding method. Within the
  uncertainties, the measured asymmetries are in good agreement with
  the SM prediction. The challenge for forthcoming measurements will
  be especially the reduction of the systematic uncertainties.

\begin{figure*}[t]
\centering
\includegraphics[scale=0.27]{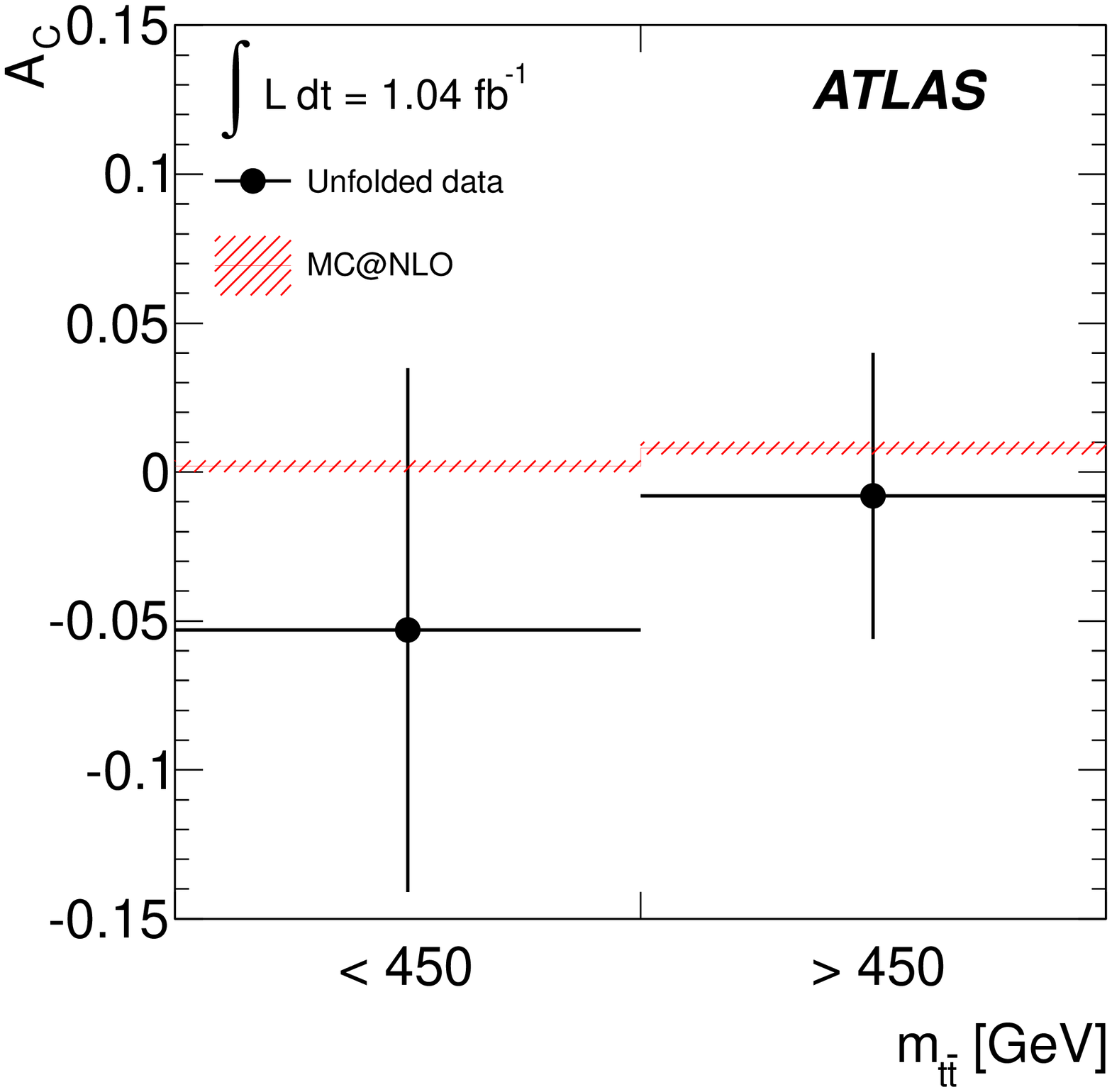}
\includegraphics[scale=0.37]{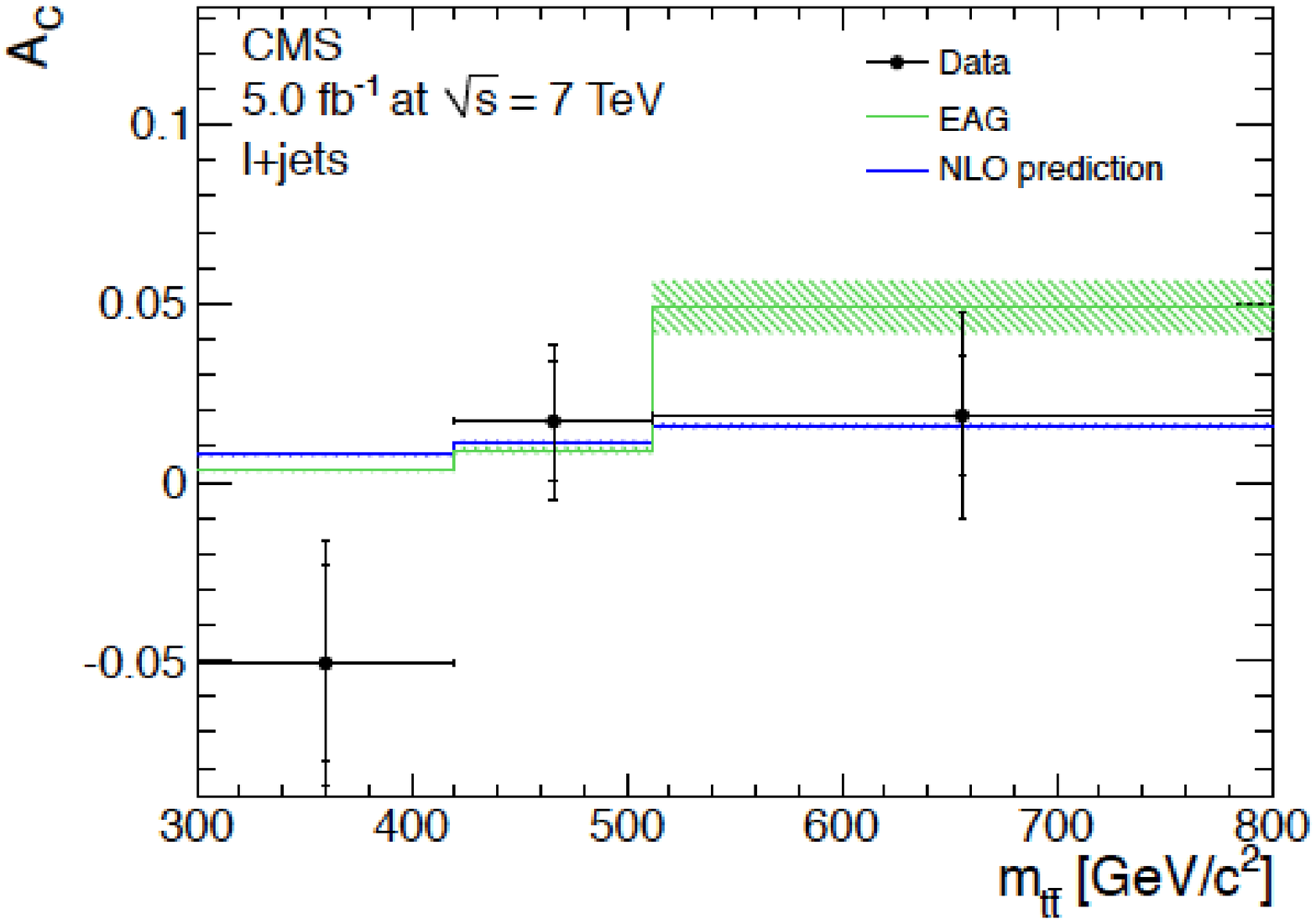}
\caption{ Parton level $A_{C}$  as  function of
  $m_{t\bar{t}}$ as extracted by the ATLAS (left) \protect
  \cite{atlasljets} and CMS (right) \protect \cite{cmsljets}
  collaborations. }  \label{lhcmtt} 
\end{figure*}

Besides the inclusive $A_C$ measurement, both collaborations also studied
 the dependency on several variables. While the fraction of
$t\bar{t}$ production via 
$q\bar{q}$ annihilation increases with larger
$m_{t\bar{t}}$, the $p_T^{t\bar{t}}$ distribution is sensitive to the
ratio of negative and positive contributions to the asymmetry. The
dependency of the asymmetry on the rapidity is caused by the effect
that gluon-gluon fusion is more dominant in the central rapidity
region, while the $q\bar{q}$ annihilation contributes more to the
forward rapidity region. The ATLAS collaboration studied the asymmetry
$A_C$ as function of $m_{t\bar{t}}$, while CMS measured the asymmetry
dependence on $m_{t\bar{t}}$, rapidity and $p_T^{t\bar{t}}$. Within
the uncertainties, no significant dependency of $A_C$ on any of the
variables under study could be noticed. CMS also compared the data
with a 
model featuring an effective axial-vector coupling of the
gluon. Figure~\ref{lhcmtt} shows the asymmetry $A_C$ as function of
$m_{t\bar{t}}$ as measured by the ATLAS (left) and CMS (right)
collaborations.

The ATLAS collaboration has performed a measurement of $A_C$ and
$A_C^{ll}$ in the dilepton final state, using 4.7~fb$^{-1}$ of 7~TeV data. For the $A_C^{ll}$
measurement, the pseudorapidity of the two leptons are used. The
reconstruction of the $t\bar{t}$ system was performed  using a matrix
element based reconstruction technique. The correction for detector
and acceptance effects was performed by using a calibration curve. The
details of this analysis can be found in Ref.~\cite{cuthdilep}. The
extracted inclusive dileptonic asymmetry is $A_C^{ll}= 2.3 \pm 1.2
{\rm (stat)} \pm 0.8 {\rm (syst)}$\%, which are compared to a SM
prediction from \mcatnlo\ of $A_C^{ll}= 0.4 \pm
0.1$\%~\cite{atlasdilep}. The charge asymmetry comes out at
$A_C=5.7\pm 2.4 {\rm (stat)} \pm 1.5 {\rm (syst)}$\%. Both results are
in good agreement with the SM prediction within the uncertainties. 
The ATLAS collaboration performed a combination of the $A_C$
measurement in the $\ell$+jets and dilepton final states, resulting in
$A_C=2.9 \pm 1.8 {\rm (stat)} \pm 1.4 {\rm (syst)}$\%, which is in good
agreement with the SM prediction. 

The ATLAS and CMS collaborations also studied the top quark
polarization~\cite{toppolatlas,toppolcms}, both measuring a value
compatible with the SM prediction. Details about these studies can be
found in Ref.~\cite{tony}.

\section{$t\bar{t}$ Asymmetries at the Tevatron and LHC}
\begin{figure*}[t]
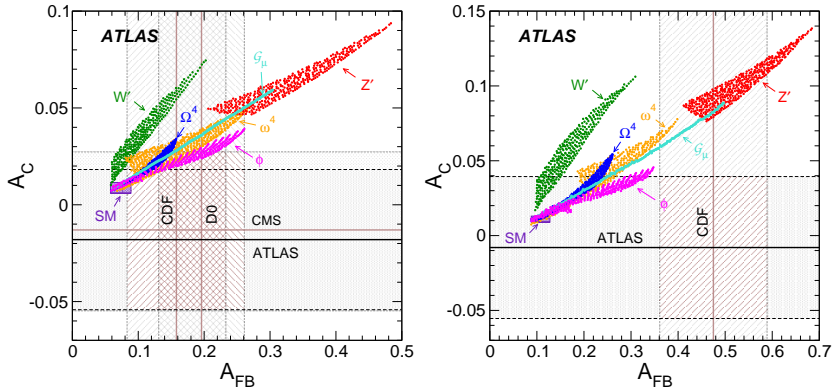

\centering
\includegraphics[scale=0.29]{tevlhc.eps}
\includegraphics[scale=0.29]{tevlhcmass.eps}
\caption{ Measured  $A_{fb}^{t\bar{t}}$ ($A_{FB}$ in the figures) from the Tevatron and charge
  asymmetries $A_C$ from the LHC, compared to predictions from the SM and
  predictions for various potential new physics
 models \protect \cite{aguilar}. The horizontal (vertical) bands and lines correspond
  to the ATLAS and CMS (CDF and D0) measurements. Left: inclusive
  asymmetry measurements. Right: Asymmetry measurements for
  $m_{t\bar{t}}>450$~GeV \protect \cite{atlasljets}. }  \label{tevatronlhc} 
\end{figure*}

While the $t\bar{t}$ asymmetries measured at the Tevatron show a
deviation with respect to the SM towards more positive values, the
measurements performed by the ATLAS and CMS collaboration of the
charge asymmetries come out to be compatible with the SM. For several
models beyond the SM, the behaviour of $A_{fb}^{t\bar{t}}$ at the
Tevatron can be different than $A_C$ at the LHC, depending, for example, on the
production process of the model under consideration. In
Fig.~\ref{tevatronlhc}, the measured asymmetries $A_{fb}^{t\bar{t}}$
from the Tevatron are plotted versus the charge asymmetry $A_C$
measured at the LHC~\cite{atlasljets}. The measurements are compared to the prediction from the
SM and various models beyond the SM that could explain a positive
asymmetry as measured at the Tevatron. With this comparison, several
of the models are disfavoured for most of their potential
parameters~\cite{aguilar}. For example, using the inclusive measurement, the $Z'$
model is in tension with the measurements, while the asymmetries for
large $m_{t\bar{t}}$ show a tension for several other models between
their prediction and the measurements.

\section{Conclusion and Outlook}
The large positive asymmetries as measured by the CDF and D0
collaboration are one of the most interesting results in the top quark
sector today. While the great performance of the LHC provided a hugh
amount of $t\bar{t}$ events, the asymmetry measurement is more challenging at the
LHC than at the Tevatron, resulting in large uncertainties on
the charge asymmetries measured by ATLAS and CMS compared to the small
SM prediction. With yet more data to be collected at ATLAS and CMS
within the next years and the progress on understanding the systematic
uncertainties, it will stay interesting to see 
whether a deviation of the SM prediction also will show up at the LHC
experiments. 


\section*{Acknowledgements} 
I thank my collaborators from ATLAS, CDF, CMS and D0 for their help in preparing the
presentation and this article. I also thank the staffs at Fermilab, CERN and
collaborating institutions, 
and acknowledge the support from the Helmholtz association.

\end{document}